\documentclass{pnastwo}

\usepackage{t1enc}
\usepackage[dvips]{graphicx}
\usepackage{pnastwof}
\usepackage{amssymb,amsfonts,amsmath}
\usepackage{float}

\def\apj{{\it Astrophys.~J.\ }}
\def\apjl{{\it Astrophys.~J.~Lett.\ }}
\def\aj{{\it Astron.~J.\ }}
\def\mnras{{\it Mon.~Not.~R.~Astron.~Soc.\ }}

\def\apjs{{\it Astrophys.~J.~Suppl.\ }}

\def\aap{{\it Astron. Astrophys.}\ }
\def\aa{{\it Astron. Astrophys.\ }}

\url{www.pnas.org/cgi/doi/10.1073/pnas.0709640104}
\copyrightyear{2013}
\issuedate{Issue Date}
\volume{Volume}
\issuenumber{Issue Number}

\begin{document}

\title{Spectra as Windows into Exoplanet Atmospheres}  

\author{Adam Burrows\affil{1}{Princeton University, Princeton, NJ 08544 USA}}

\contributor{Accepted to Proceedings of the National Academy of Sciences}

\maketitle

\begin{article}

\begin{abstract}

Understanding a planet's atmosphere is a necessary
condition for understanding not only the planet itself,
but also its formation, structure, evolution, and habitability,
This puts a premium on obtaining spectra, and developing 
credible interpretative tools with which to retrieve vital planetary information.
However, for exoplanets these twin goals are far from being realized.
In this paper, I provide a personal perspective on exoplanet theory 
and remote sensing via photometry and low-resolution spectroscopy.
Though not a review in any sense, this paper highlights the limitations in our 
knowledge of compositions, thermal profiles, and the effects of stellar irradiation,
focussing on, but not restricted to, transiting giant planets.
I suggest that the true function of the recent past of exoplanet atmospheric 
research has been not to constrain planet properties for all time, but to
train a new generation of scientists that, by rapid trial and error,
is fast establishing a solid future foundation for a robust science of exoplanets.

\end{abstract}


\section{Introduction}
\label{intro}

The study of exoplanets has exponentiated since 1995, a trend that 
in the short term shows no signs of abating.  Astronomers 
have discovered and provisionally studied more than a hundred {\em times} 
more planets outside the solar system than in it.  Statistical and orbital 
distributions of planets across their broad mass and radius continuum, 
including terrestrial planets/Earths, ``super-Earths," ``Neptunes," and giants, are emerging
at a rapid pace.  

However, understanding its atmosphere is a necessary 
condition for understanding not only the planet itself, 
but also its formation, evolution, and (where relevant) habitability, 
and this goal is far from being realized.  
Despite multiple ground- and space-based campaigns to characterize 
their thermal, compositional, and circulation patterns (mostly for transiting giant planets), 
the data gleaned to date have (with very few exceptions) been of marginal utility.  
The reason for this is that most of the data are low-resolution photometry at a few broad bands 
that retain major systematic uncertainties and large error bars. Moreover, the
theory of their atmospheres has yet to converge to a robust and credible
interpretive tool. The upshot of imperfect theory in support of imprecise data
has been ambiguity and, at times, dubious retrievals.  To be fair, i) telescope assets
are being employed with great effort at (and, sometimes, beyond) the limits of their designs; 
and ii) most planet/star contrast ratios are dauntingly small.
As a consequence, the number of hard facts obtained over the last ten years 
concerning exoplanet atmospheres is small and by no means commensurate with the 
effort expended.

An important aspect of exoplanets that makes their characterization an 
extraordinary challenge is that planets are not stars.  They have character and greater complexity.
A star's major properties are determined once its mass and metallicity are known.
Most stars have atmospheres of atoms and their ions.  However, planets have molecular 
atmospheres with elemental compositions that bespeak their formation, accretional, 
and (where apt) geophysical histories.  Anisotropic stellar irradiation, clouds, and rotation
can break planetary symmetry severely, with the clouds themselves introducing multiple 
degrees of complexity, still unresolved even for our Earth. Molecules have much more complicated spectra
than atoms, with a hundred to a thousand of times more lines, and irradiated objects
experience complicated photochemistry in their upper reaches. It took stellar atmospheres $\sim$100
years to evolve as a discipline, and it still is challenged by uncertainties
in oscillator strengths and issues with Boltzmann and thermal equilibrium. 
Furthermore, the spectroscopic databases for molecules \cite{HITRAN}, particularly at the high temperatures 
(500-3500 K) experienced by close-in transiting planets, are much more incomplete than those
for atoms, and the relevant collisional excitation rates are all but non-existent.
Therefore, it can reasonably be suggested that the necessary theory for detailed studies
of exoplanets is in its early infancy.

One might have thought that the study of our solar system had prepared us 
for exoplanetology, and this is in part true.  The solar system has been a great, perhaps
necessary, teacher.  However, most solar-system spectra are angularly-resolved with a long time baseline
and high signal-to-noise.  Exoplanets will be point sources for the foreseeable future, and 
signal-to-noise will remain an issue. Perhaps more importantly, much solar-system research is 
conducted by probes in-situ or in close orbit, with an array of instruments for direct determination
of, for example, composition, surface morphologies, B-fields, charged-particle environments, 
and gravitational moments. Masses and radii can be exquisitely measured.  
Orbits are known to standard-setting precision. Moreover, when comparing measured with theoretical spectra,
the latter are often informed by direct compositional knowledge. 

This is not the scientific landscape that we can envision for exoplanets. Exoplanet science is
an observational science that must rely on the astronomical tools of remote spectroscopic 
sensing to infer the physical properties of individual planets.  This puts a premium on
obtaining spectra, and developing interpretative toolkits in the tradition of classical
astronomy, without the luxury of direct, in-situ probes.  Hence, though solar-system
variety will continue to inform exoplanet thinking and motivate many calculations, 
the methodology of solar-system research is not the best model for conducting exoplanet 
research.  Rather, we must determine the most robust and informative methods with which to
interpret remote spectra and perform credible spectral retrievals of physical properties. Hence, the
science of exoplanet characterization is better viewed as a science of spectral diagnostics, and 
developing this art should be our future focus.

To date, planets that transit their stars due to the chance orientation of their orbit planes have
provided some of the best constraints on hot exoplanet atmospheres.  The variation of transit 
depth\footnote{equal to the planet/star area ratio, $\left(\frac{R_p}{R_*}\right)^2$, where $R_p$ and
$R_*$ are the planet and star radii, respectively $-$ for giants, $\sim$1\%; for Earths $\sim$0.01\%.} 
and, hence apparent planet radius (R$_p$), with wavelength ($\lambda$) is an ersatz spectrum 
and can be used to infer the presence of chemical species with the corresponding  
cross-sections. Water, sodium, and potassium have been 
unambiguously detected by this means.  Approximately $180^{\circ}$ out of phase with the primary transit,  
when the same planet is eclipsed by its star, the difference between the summed light of planet and star and 
that of the star alone reveals the planet's light.  This is the secondary eclipse, and such measurements, 
when performed as a function of wavelength, render the planet's emission spectrum; measurements
taken between primary and secondary eclipse provide phase light curves. The secondary eclipse
planet/star flux ratio is lower than the transit depth by approximately $\sim$$\left(\frac{R_*}{2a}\right)^{1/2}$, 
where $a$ is the orbital semi-major axis and R$_*$ is the stellar radius. This can be a factor of one tenth. 


The transit and radial-velocity techniques with which most exoplanets have been found
select for those in tight orbits.  Tight orbits at the distances of stars in the solar neighborhood
subtend very small angles (micro-arseconds to 10's of milli-arcseconds), and such angular proximity to a bright 
primary star mitigates against direct planet detection, imaging, or characterization.  
For wider separations of tens of milliarseconds to arcseconds, the 
resulting contrast ratios for terrestrial and giant planets in the optical of $10^{-10} - 10^{-6}$,  
and in the near- to mid-infrared of $\sim$$10^{-8} - 10^{-4}$, are quite challenging\cite{burrows_2005}.  
However, such direct planet imaging is not only now conceivable, but has 
been accomplished. Four giant exoplanets around HR 8799 \cite{marois,hr8799_madhu} and one around 
$\beta$ Pictoris \cite{betapic}, with masses of $\sim$5 $-$ 15 Jupiter masses (M$_J$) and angular separations 
between $\sim$0.3 and $\sim$1.5 arcseconds, have recently been found. 
As direct imaging techniques mature, more and smaller directly-imaged planets will be discovered.
However, as articulated earlier, it is only with well-calibrated spectral measurements at useful resolutions 
that we can hope to characterize wide-separation exoplanet atmospheres robustly. 
Polarization measurements will also have an important diagnostic role, particularly for cloudy atmospheres,
which at quadrature should be polarized in the optical to tens of percent. 

Currently, due to their larger size, the photometric and spectroscopic techniques mentioned 
above have been applied mostly to giant exoplanets. Earths are ten times smaller in radius 
and one hundred times smaller in mass.  Hence, while astronomers and theorists
hone their skills on the giant exoplanets, fascinating in their own right, these giants are also serving as 
stepping stones to the smaller planets, in anticipation of future routine campaigns to characterize 
them as well.  Therefore, I concentrate in this article on the giant population, but all of the 
basic methodologies employed in their study can be translated bodily to the investigation of  
smaller ``exo-Neptunes," terrestrial planets/Earths, and ``super-Earths."

A comprehensive review would necessitate more pages than this more synoptic and summary opinion piece can provide, 
but for those readers interested in an expanded treatment there are numerous archival papers
from which to draw. They cover topics such as the general theory of giant exoplanets \cite{burrows_1997},
giant planet atmospheres \cite{burrows_orton,burrows_rmp_2001,sudarsky_2003}, analytic atmosphere theory \cite{hubeny_2003,guillot_2010},
opacities \cite{freedman,sharp_burrows,HITRAN}, thermochemistry and elemental abundances 
\cite{lodders_2003,lodders_fegley,lodders_fegley_companion,burrows_sharp}, the chemistry of hot Earth 
atmospheres \cite{fegley_silicate_atmos}, albedos \cite{sudarsky_albedo,analytic_albedos,marley_1999,hd209_albedo_burrows},
giant planet models at wide separations \cite{burrows_2004,hr8799_madhu,burrows_2005}, phase functions \cite{sudarsky_phase,barman_2005},
irradiated atmospheres and inversions \cite{hubeny_2003}, basic transit theory \cite{seager_sasselov,brown_transit},
transit spectra for Earths \cite{kalten_2009}, emission spectra of Earth-like planets \cite{kalten_2007},
transit spectra of Earth-like planets \cite{ehrenreich}, habitable zones \cite{kasting_1993}, 
theoretical exo-Neptune spectra \cite{burrows_spiegel_neptunes}, planet polarization \cite{Seager_polar,analytic_albedos},
and clouds and hazes \cite{ackerman,marley_2013,helling}.  In this paper, I refer preferentially to my own work,
but suggest that the general conclusions arrived at here have broad applicability.


\begin{figure} 
\null\hskip-1.0cm
\includegraphics[height=0.38\textheight,angle=-90]{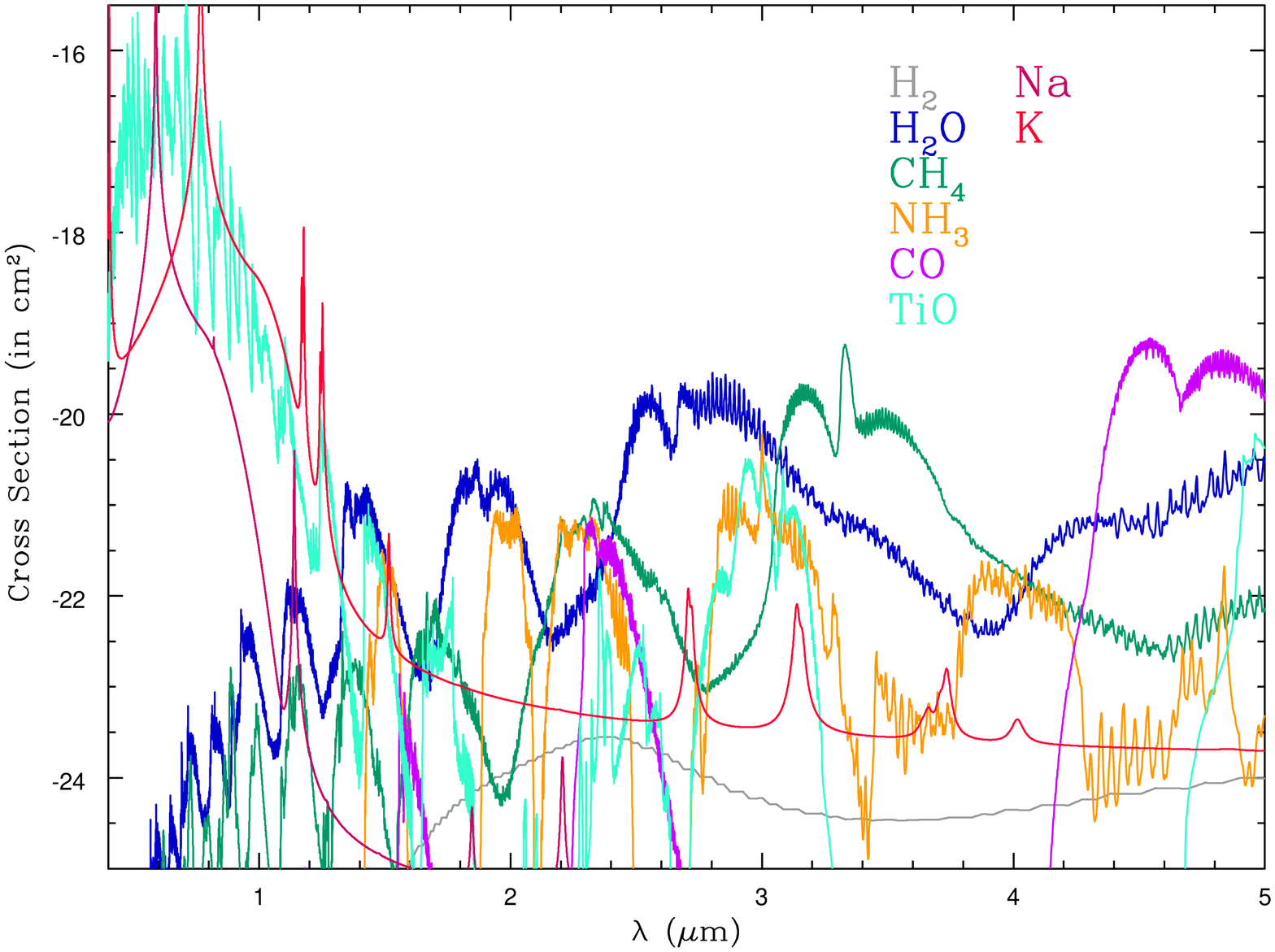}
\null\hskip-0.5cm
\includegraphics[height=0.38\textheight,angle=-90]{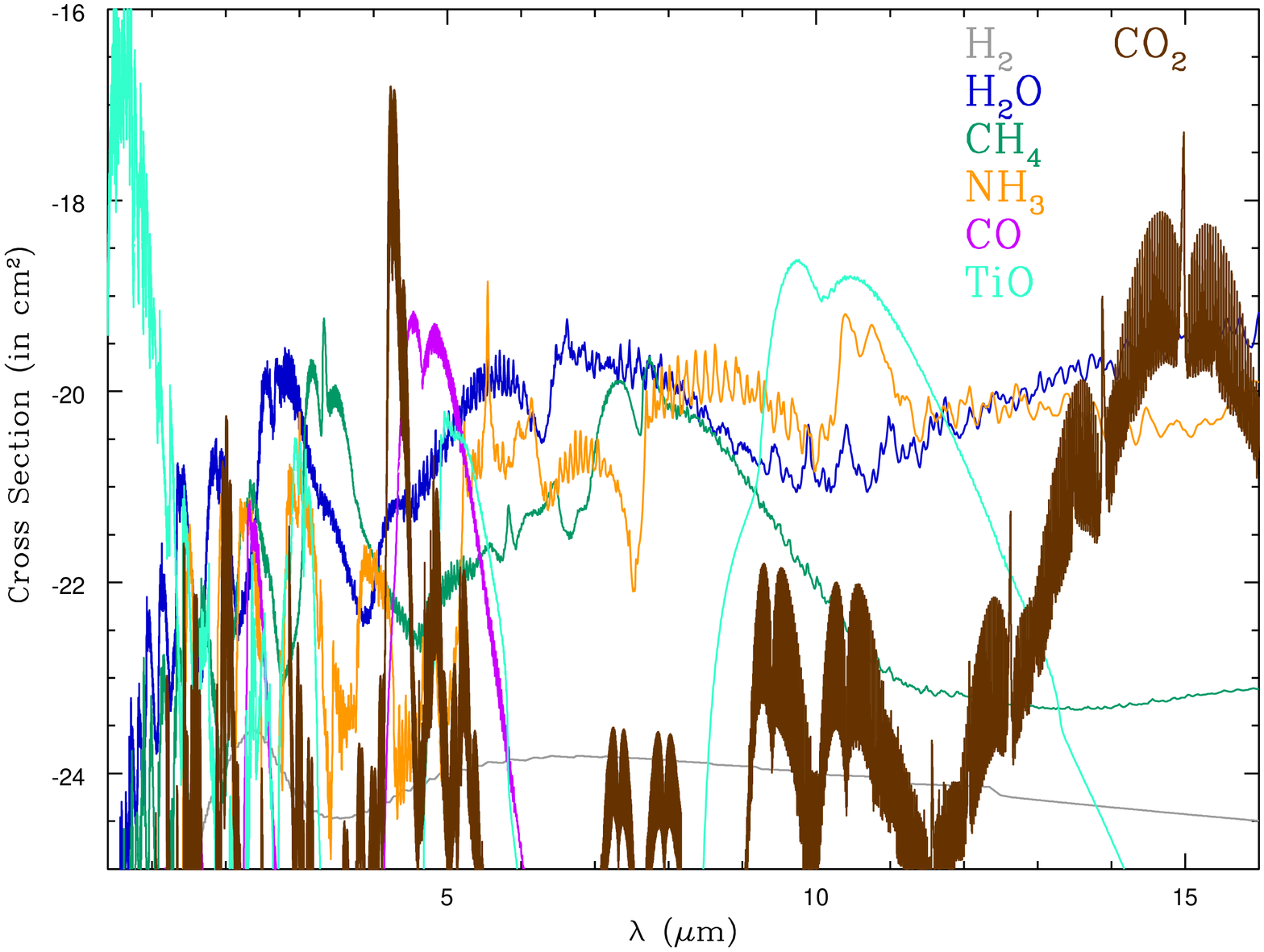}
\caption{{\bf Top:} The figure on the top depicts the logarithm base 10 of the cross section
per molecule or atom (in cm$^2$) versus wavelength (in microns) from 0.4 to 5.0 $\mu$m for
various important species thought to be prominent in the atmospheres of exoplanets,
in particular giant exoplanets.  They are H$_2$ (gray), H$_2$O (blue), CH$_4$ (green), NH$_3$ (orange), TiO (cyan),
Na (red; leftmost, with strong peak at 0.589 $\mu$m), and K (red; rightmost, with strong peak at 0.77 $\mu$m).
Other molecules of note (not depicted) are CO$_2$, N$_2$O, O$_2$, and O$_3$.  For presentation purposes, these cross sections
have been calculated at 1500 K and 100 bars.  The latter is far too high a pressure to be representative of
regions in exoplanet atmospheres that can be probed, but was employed to more clearly distinguish individual features.  Importantly,
the wavelengths of the major bands and lines are not significantly temperature- or pressure-dependent, though
their strengths are. {\bf Bottom:} The figure on the bottom is the same plot, but extended to 16 $\mu$m to highlight the mid-infrared
and to include CO$_2$ (brown) at 296 K and atmospheric pressure \cite{kratz}. Note the prominent CO$_2$ feature at $\sim$15 $\mu$m.
The spectral features for each chemical species are crucial discriminating diagnostics for remote exoplanetary
sensing and characterization.  See text for a discussion.
\label{fig1}}
\end{figure}

\section{Compositions and Opacities}
\label{opac}

The variety of compositions found in the gaseous atmospheres of solar-system planets
suggests that that for exoplanet atmospheres must be at least as broad.  Generally 
lower in temperature than stellar atmospheres, planetary atmospheres are 
dominated by molecules.  Though fractionation and differentiation processes are
no doubt involved in their formation, their elemental abundances should 
reflect the most abundant elements in the Universe.  For giant exoplanets (like
brown dwarfs\footnote{It is likely that the brown dwarf and giant planet mass functions overlap,
so that a tentative assignment is generally premature. A flexible and open-minded philosophy 
towards nomenclature is then best\cite{burrows_rmp_2001}, which more data will progressively guide
towards a more reasonable classification scheme. I do note, however, that much recent 
data for giant planets has been for the close-in transiting subset.  For these, the 
fact that these are irradiated, while free-floating brown dwarfs are not, 
significantly alters the colors and atmospheric characteristics of the former, 
when they might otherwise have had spectra like isolated low-mass brown dwarfs 
(see figures in the Supplement). One can speculate that, barring the irradiation difference, differences in 
atmospheric abundances, rotation rates, and orbital regimes might eventually distinguish brown dwarfs 
from giant planets (at least statistically).}), this means H$_2$, He, H$_2$O, CO, CH$_4$, NH$_3$, PH$_3$, H$_2$S, 
Na, K predominate, with most of the metals sequestered in refractories at 
depths not easily penetrated spectroscopically. However, titanium and vanadium oxides (TiO and VO),
identified in cool-star and hot-brown-dwarf atmospheres, have been suggested to reside in quantity 
in the upper atmospheres of some hot Jupiters to heat them by absorption in the optical 
and create inversions\cite{fortney_2008}.  However, TiO and VO too are likely condensed out \cite{spiegel_tio}. Since such 
inversions require an optical absorber at altitude, what this absorber is, molecule 
or absorbing haze/cloud, remains a major mystery\cite{burrows_2007_209}.

For terrestrial planets, the molecules N$_2$, CO$_2$, O$_2$, O$_3$, N$_2$O, and HNO$_3$ must 
be added to the list above, with O$_2$, O$_3$ (ozone), and N$_2$O considered biosignatures, 
along with the ``chlorophyll red edge" (or its generalization). Many other compounds 
could be envisioned, and there is added complexity to terrestrial planet atmospheres 
due to atmosphere-surface interactions that are so important, for example, for our Earth.  
The major constituents of ``Neptune" atmospheres are likely closer to those of giants, but the relative
abundances in any exoplanet atmosphere must be considered as yet poorly constrained. 
Constraining these abundances is a goal, however, and one does so by identifying 
their unique signatures in measured atmospheric spectra and comparing the observed 
spectrum in its totality with spectral models.  This extraction is ``retrieval,"
which at a minimum should also yield temperatures and temperature profiles.  Since many
parameters characterize exoplanet atmospheres (e.g., species, abundances, 
temperatures, spatial distributions, gravities, haze and cloud layers), 
the low information content of few-band photometry is not adequate to avoid the pitfalls
of parameter degeneracy.  This, however, with very few exceptions, is the 
current situation in exoplanet research.  With too few data points in pursuit of too many quantities,
interpretation is thereby severely compromised and error-prone.  It is only with good-resolution spectra, 
with small and credible error bars, that we can establish robust conclusions about exoplanets
and build a solid future for the subject. This is a challenge, but a necessity.
 
Helium and N$_2$ have weak spectral features.  A prominent O$_2$ feature is the Fraunhofer A-band
at 0.76 $\mu$m, and the signal feature for O$_3$ is the band at 9.6 $\mu$m.  Rayleigh scattering off molecules 
roughly follows a $\lambda^{-4}$ dependence, is proportional to the summed product of molecular polarizability 
and abundance, and is most relevant only in the blue and UV in reflection. 

Figure \ref{fig1} depicts example gas-phase absorption cross sections per molecule (or atom) 
versus wavelength \cite{sharp_burrows,HITRAN} for H$_2$, H$_2$O, CH$_4$, CO, Na, K, and CO$_2$ \cite{kratz}.  These species are 
expected to be important in giant exoplanet atmospheres (for which we currently have the most data), 
but are also likely important (to varying degrees) in terrestrial, super-Earth, and exo-Neptune atmospheres.
In the top plot, we focus on the 1.0 $-$ 5.0 $\mu$m range and include the TiO, Na, and K opacities so prominant in the optical, 
while the bottom plot extends to 15 $\mu$m to reveal the behavior in the mid-infrared and the signature feature 
of CO$_2$ at $\sim$15 $\mu$m. 

As indicated in Figure \ref{fig1}, strong water features are ubiquitous and are found 
at (roughly) 0.94, 1.0, 1.2, 1.4, 1.9, 2.6, and 5 $-$ 7 microns, defining between them the $I$, $Z$, $J$, $H$, $K$, 
and $M$ bands through which much of ground-based near-infrared astronomy is conducted.
Methane has important features at 0.89, 1.0, 1.17, 1.4, 1.7, 2.2, 3.3, and 7.8 microns.
Carbon monoxide stands out at 2.3 and 4.5 microns, while CO$_2$ has diagnostic features near 2.1, 4.3, and 15 microns.
Ammonia has many features, but the one at 10.5 microns is most noteworthy.  Molecular hydrogen (H$_2$) has 
no permanent dipole, but one can be induced by collisions (``collision-induced absorption") 
at high pressure, and the result is a family of undulations from $\sim$2.2 to $\sim$20 microns
that has been seen in Jupiter, Saturn, and brown dwarfs.  A central goal of transit, 
reflection, or emission spectroscopy of exoplanets is to identify these species (and perhaps infer their abundances)
by these distinctive features.

\section{Clouds and Hazes}
\label{clouds}

Condensates can form and reside in exoplanet atmospheres as clouds or hazes \cite{marley_1999,marley_2013} and
can have a disproportionate influence on spectra.  This is because, assembled in a grain, 
such aggregations can respond coherently to light (depending upon the particle size and wavelength). 
So, very little areal mass density can translate into a large optical depth
and a trace species can loom large. In addition, with a spectrum of particle sizes 
and enhanced line broadening in the grain, their absorption and scattering cross sections 
can have a continuum character and veil a wide spectral range.  The result can be partial 
(or complete) muting of the gas-phase spectral features, making understanding 
condensates and incorporating their effects into models as important as it is difficult.  
To properly handle the effects of clouds we need to know the condensate species, 
grain size and shape distributions, the complex index of refraction, and the spatial distribution 
in the atmosphere.  Such knowledge is generally in short supply.

The possibility of water clouds in terrestrial atmospheres is uncontroversial, the presence of 
ammonia clouds in the atmospheres of Jupiter and Saturn has been observed in detail, and 
the central role of silicate and iron clouds in brown dwarf L dwarfs is reasonably inferred
by their very red infrared colors. These situations are in part informed by known thermochemistry.  
However, water clouds are expected in cold giant exoplanet and brown dwarf atmospheres \cite{sudarsky_phase,beyond_tdwarfs};
Na$_2$S and KCl clouds are thought to reside in late T dwarf brown dwarfs; an extra absorber 
in the optical and at altitude has been invoked to explain the inversions and over-hot atmospheres 
inferred from the spectra at secondary eclipse of some transiting hot Jupiters\cite{burrows_2007_209}; 
a thick haze envelopes the atmosphere of Saturn's Titan; and there is a trace absorber in the blue 
that makes Jupiter and Saturn redder than Neptune or Uranus. None of the causative species
in these situations is either known, or if known, well-modeled.  The case of Jupiter's color 
is a cautionary tale. The factor of two suppresion in its reflected blue flux could be due to traces 
at the part in $10^{10}$ level of either polyacetylenes, sulfur or phorphorus compounds, tholins, or something else \cite{sudarsky_albedo}.  
Such leverage by a small (and unknown) ``actor" in the interpreation of such a large effect should give one pause,
and emphasizes the potential complexity of the task of exoplanet characterization. 
Photolytic chemistry is likely a cause in Titan's atmosphere, as in many other contexts, 
but this is small comfort when designing a modeling effort aimed at anticipating all reasonable 
possibilities.

Scattering in general is important only in reflection and transit spectra, not in emission, 
and is most prominent for hazes and clouds. In fact, longward of the ultraviolet (UV),
clouds are necessary to give a planet any appreciable reflection albedo above $\sim$1\% \cite{sudarsky_albedo}.
Also, in reflection, as a general rule, cloud or 
UV/blue Rayleigh scattering can yield highly polarized fluxes \cite{analytic_albedos}. The polarized fraction is 
higher when the absorption fraction is higher and the scattering albedo\footnote{the ratio of 
the scattering cross section to the total cross section} is low, 
but in this case the overall reflected flux is low.  This suggests that polarization might 
in some circumstances be a useful ancillary diagnostic of exoplanet atmospheres. Unlike for 
gas species, for many realizations of likely hazes or clouds in exoplanet atmospheres, the scattering albedo
can be either high or low, depending upon species and wavelength range, and is frequently high.
This suggests that reflection spectra can be dominated by the effects of such layers,
and, moreover, that transit spectra can be affected by particulate scattering (as opposed to only absorption).
Clearly, one must be aware of the possible presence of clouds and hazes when performing exoplanet 
spectral retrievals.

\begin{figure} [h!]
\null\hskip-0.5cm
\includegraphics[height=0.35\textheight,angle=-90]{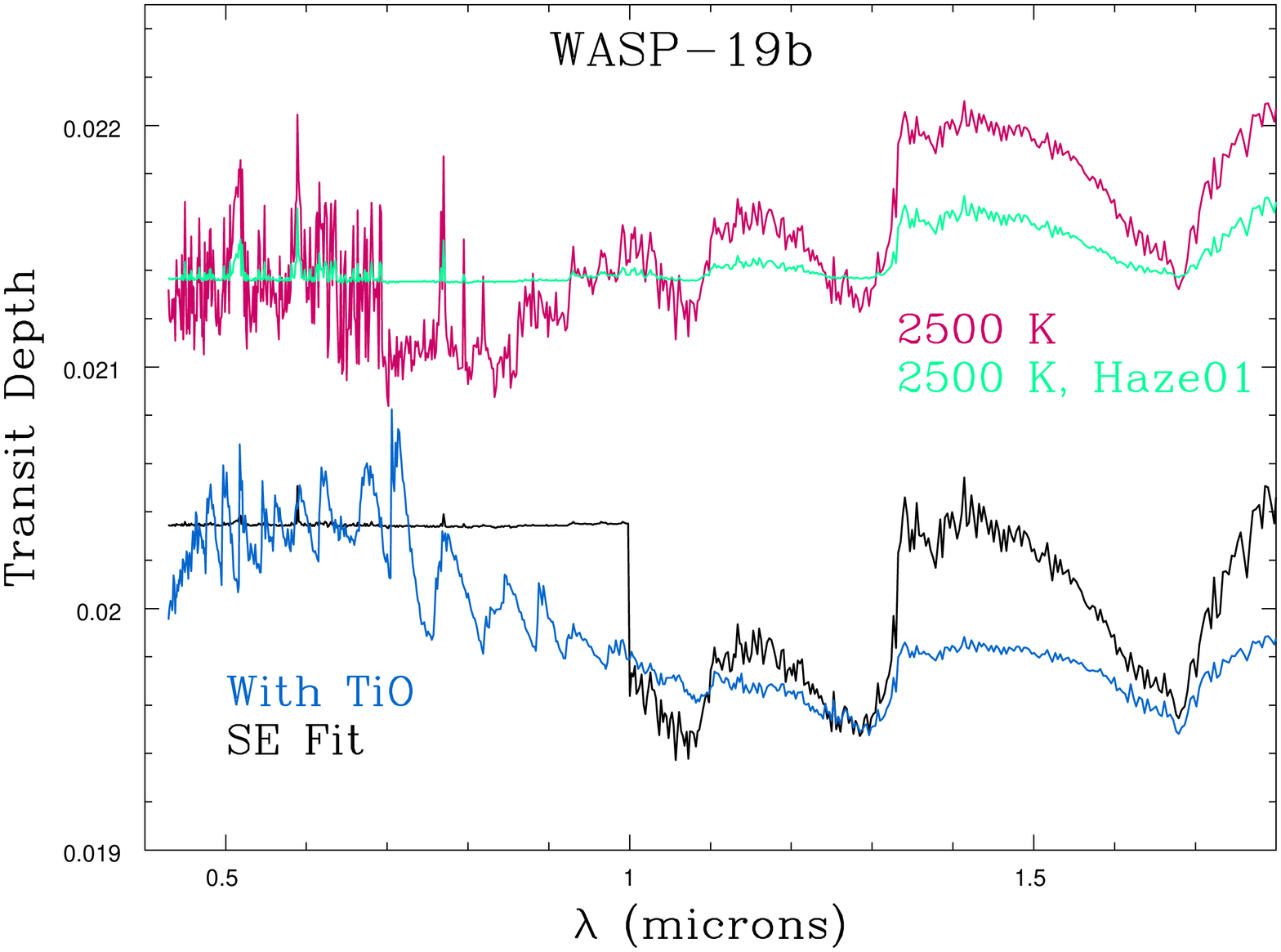}
\null\hskip-0.0cm
\includegraphics[height=0.35\textheight,angle=-90]{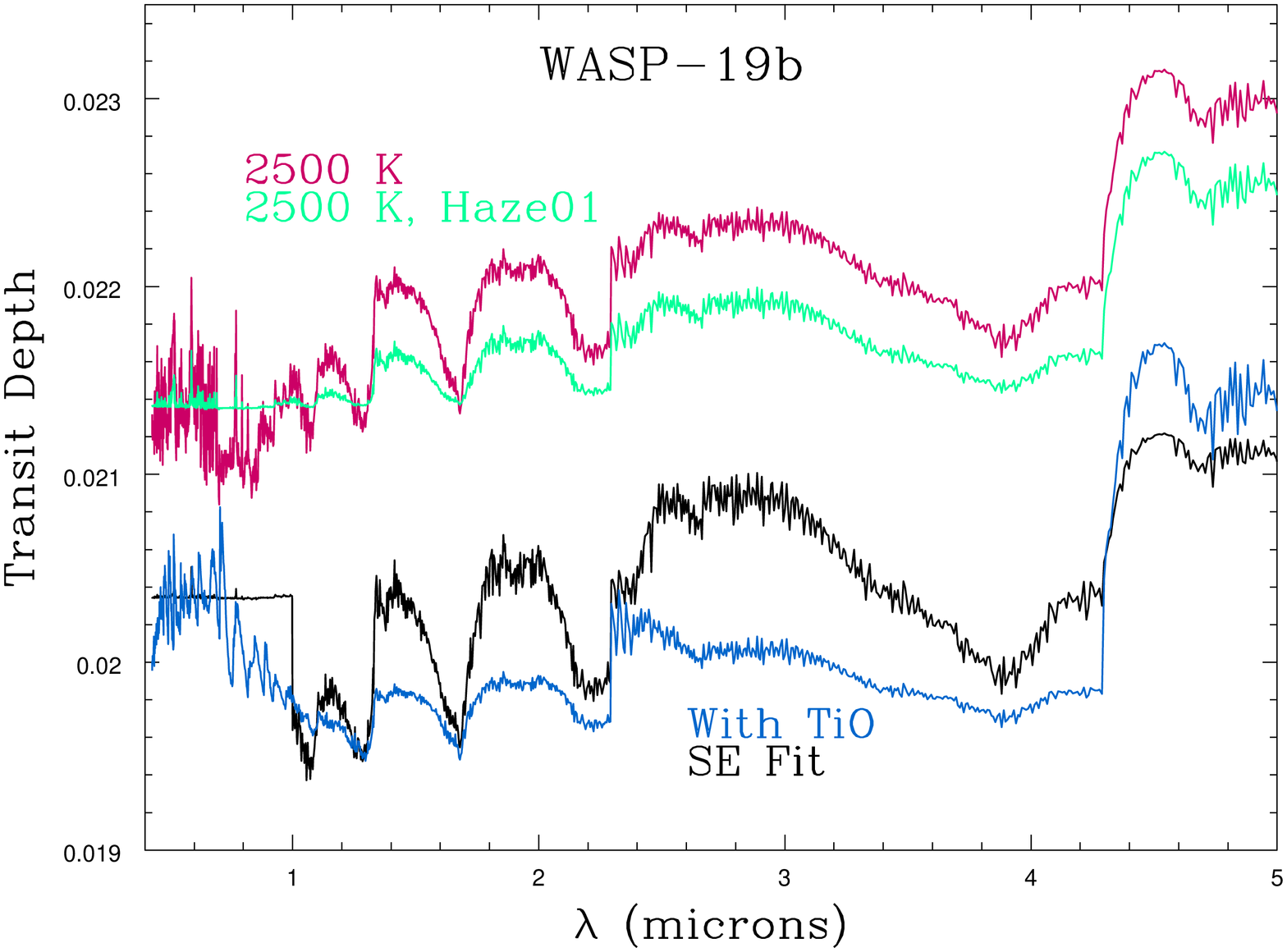}
\caption{{\bf Top:} Shown are model fractional transit depths versus wavelength (in microns) between 0.4 and 1.8 $\mu$m
for a WASP-19b-like planet. The blue curve is a dayside model with TiO in its atmosphere and a redistribution parameter,
P$_n$, of 0.3\cite{burrows_2007_209}, that is irradiated by a stellar model of WASP-19 at the distance of WASP-19b. The black curve (``SE Fit") is
a dayside model with a P$_n$ = 0.3 and an ``extra absorber" at altitude with an opacity of 0.05 cm$^2$ g$^{-1}$ from 0.4 to 1.0 $\mu$m,
configured to fit the measured Spitzer/IRAC secondary eclipse data. The red and green models have isothermal atmospheres at 2500 K, 
with the flatter green model having a uniform haze with an opacity of 0.01 cm$^2$ g$^{-1}$. {\bf Bottom:} The bottom plot is the same, but extended
to 5.0 microns.  In all models shown, water features (see Figures \ref{fig1}) are the most prominent, while TiO features are in
evidence in the TiO model and the effect of a veiling haze is manifest in that model. Note that for this exoplanet
the magnitude of the variation with wavelength is generally less than or equal to a part in a thousand. See text for details. \label{fig2}}
\end{figure}

\section{Transit Spectra}
\label{transit}

Transit spectra\footnote{Often referred to imprecisely as ``transmission spectra." What 
one is actually measuring is the transit depth, which reflects what is {\em not} transmitted.  
In addition, the implication of the term ``transmission" is that we are imaging the planet's limb region
and measuring the variation in $\tau$ or $e^{-{\tau}}$.  However, we are actually probing
1 - $e^{-{\tau}}$, its complement.} are direct probes of atmospheric scale heights and atmospheric 
abundances near the terminator(s).  However, if the atmosphere is optically thick and overlays 
a rocky core there is no obvious way to determine the core's contribution to the measured radius.
Therefore, it is standard practice to analyze transit spectra with respect to an arbitrarily 
determined reference radius, often taken to be the inferred discovery radius in the optical.
When the solid surface of a terrestrial or super-Earth planet is not a priori known, or
is inaccessible by measurement, then there will be ambiguity with respect to its contribution
to the transit depth.  This will not be the case with an airless planet, and is moot for 
a gaseous planet, but is an issue to consider when falsifying theory.

The measured fractional diminution in the stellar light at a given wavelength is the transit 
depth \cite{brown_transit,fortney_2003}.  The stellar beams pointed at the Earth probe the planet's atmosphere transversely 
along a chord perpendicular to the impact radius.  Hence, the relevant optical depth, $\tau$,
is not the depth in the radial direction associated with emission, but much larger. The 
contribution of the annulus, or partial annulus in the case of the ingress or egress phases, to the blocking
of stellar light is 1 - $e^{-{\tau}}$ times the annular area.  The sum of such terms over the entire 
atmosphere provides the integrated blocking fraction due to the atmosphere. That this $\tau$
is larger than the radial $\tau$ allows transit depth to be more sensitive to trace chemical species
than emission or secondary eclipse spectra and amplifies their effect.  This may be particularly true of
atmospheric hazes that may be too thin in the radial direction to affect emission, but are thick
along the chord\cite{howe_gj1214b}, and may be why Pont et al. \cite{pont} see an almost featureless 
transit spectrum for HD 189733b and infer a veiling haze, while the associated IRAC and IRS data at secondary 
eclipse clearly reveal water signatures\cite{grillmair}.  Another reason may be that since transit spectra
probe the terminator, the transition region between day and night, a condensate is more likely to form 
as the temperature transitions to lower values.  Be that as it may, the terminator is a complicated
region that introduces special challenges for the theory of transit spectra.

Despite this, a simple analytic model\cite{howe_gj1214b,lecavelier,burrows_rmp_2001} can be developed that captures 
the basic elements of general transit theory.   Integrating along a chord at a given impact parameter
and assuming an exponential atmosphere with a pressure scale height, $H$\footnote{$H = kT/\mu g$, where $g$ is the gravity,
$\mu$ is the mean molecular weight, $T$ is an average atmospheric temperature, and $k$ is Boltzmann's constant.}, yields an approximate 
amplification factor for the chord optical depth ($\tau_{\rm chord}$) over the radial optical depth of $\sqrt{2\pi R_p/H}$, which
can be $5-10$.  This means that the $\tau_{\rm chord} = 2/3$ condition that approximately defines the apparent
planet radius at a given wavelength is pushed to larger impact parameters (radii) and that the fractional transit depth 
is increased by a factor $\propto 2H/R_p$.  Moreover, it is straightforward to show that
\begin{equation}
\frac{dR_{p}}{d\ln{\lambda}} \approx H\frac{d\ln{\sigma}}{d\ln{\lambda}} \, ,
\label{sigma_r}
\end{equation}
where $\sigma$ is the total species-weighted interaction cross section (the sum of absorption and scattering).
Note that, whereas emission spectra (ignoring reflection) depend upon only absorption,
transit spectra depend upon both scattering and absorption processes.  In fact, the haze inferred
for HD 189733b could be purely scattering, and as such would make no contribution to the emission
at secondary eclipse.  However, it is likely that any haze has a non-unity scattering fraction/albedo,
introducing flexibility, but also further complexity, into the simultaneous interpretation of 
transit and emission spectra.

Equation \ref{sigma_r} suggests that significant wavelength variations in cross section, as across an absorption
band, translate into a change in the apparent radius of order $H$.  This is the essence of the use of
transit measurements as a function of wavelength to determine compositions.  Since $R_{p}$ depends upon the logarithm of $\sigma$,
eq. \ref{sigma_r} also indicates that the dependence upon abundance is logarithmic and, hence, weak. While it is ``easy"
to discern a molecular feature, it is not easy with transit spectra to determine its abundance. Note that 
since $H = kT/\mu g$, a low (high) temperature, high (low) gravity, or high (low) mean molecular weight atmosphere will
yield weaker (stronger) indications of composition. Hence, as long as spectroscopically 
interesting species reside in the atmosphere in reasonable abundances, a hot, H$_2$-rich atmosphere 
(without a veiling haze/cloud) yields the largest, most diagnostic, radius variations with wavelength. 

If there are differences in the compositions and scale heights at the east and west limbs of a planet,
such differences are in principle discernible as differences in ingress and egress transit spectra.
Though difficult even for a giant exoplanet, such measurements might be doable in the future and could shed light
on atmospheric dynamics and any pronounced zonal flow asymmeties.  

In addition, narrow-band, very-high-resolution
spectroscopy before and during transit has great potential to reveal planetary orbital, 
spin, and wind speeds, as well as compositions (cf. the measurement of CO lines by 
\cite{hd209_snellen_winds}).  Though giant exoplanets are the most studied population to date via 
multi-band transit photometry and spectrophotometry (as opposed to wide single-band 
observations \`a la {\it Kepler} \cite{borucki_2010}), such data around small M dwarfs for terrestrial planets 
and super-Earths (such as GJ 1214b $-$ see \cite{howe_gj1214b}, and references therein) have great promise to probe the 
atmospheres of these smaller, but likely more numerous, planets.  Measuring the emission spectra of
Earths around solar-like stars will be much more challenging.

Figures \ref{fig2} portray the general character of representative theoretical 
exoplanet transit spectra from 0.4 to 5.0 microns.  The models are for the giant WASP-19b and
include isothermal atmospheres at $T=2500$ K, with and without a uniform gray haze with an opacity 
of 0.01 cm$^2$ g$^{-1}$, a model that attempts to fit its IRAC data at secondary eclipse \cite{anderson} 
with an unknown ``extra absorber" at altitude of constant optical opacity 0.05 cm$^2$ g$^{-1}$ 
(from 0.4 to 1.0 micron), and a similar model employing TiO as the extra absorber. For clarity, 
the latter two are shifted arbitrarily from the former two.  We note that the transit depth is
of order $\sim$2\% and that the variation due to the presence of water bands is approximately 
one part in a thousand. The depths for other hot Jupiters could vary with wavelength by as little 
as a few parts in ten thousand.

One sees immediately that the extra optical absorber, whatever its nature, increases the ratio of the optical to
infrared radii, that the TiO hypothesis can readily be falsified, that the spectral features
of (here) water should be readily detected\footnote{In fact, water has already been detected in several
giant planet atmospheres via transit spectra (e.g., \cite{deming_haze}).}, that the radius variations in the mid-infrared can
be of larger amplitude, and that even low-opacity hazes can mute these variations substantially.
The diagnostic potential of transit spectra is manifest in plots such as these.
It is equally clear that the interpretation of but a few photometric points with significant
error bars are ambiguous.  Good {\it spectra} are the key.

\section{Secondary Eclipse}
\label{second}

For a circular orbit, when 180$^{\circ}$ out of phase with the transit, 
the planet is occulted by the star and is in secondary eclipse. During the 
eclipse, the summed light of the planet and star being monitored 
shifts to that of the star alone, and by the difference 
the planet's emissions are determined.  The {\it Spitzer} space telescope \cite{werner} has been
particularly productive in this mode, providing near- and mid-infrared 
photometric points for $\sim$30 nearby transiting planets (mostly giants).
For close-in planets, for which the transit probability is largest, the planet 
is emitting mostly reprocessed stellar light \cite{burrows_2006_secondary,burrows_2008_close-in}. Stellar irradiation and zonal 
atmospheric winds and dynamics break the simple spherical symmetry, so that 
3D models would seem most appropriate.  However, such models have yet to prove 
themselves and simpler 1D hemisphere-averaged models have been employed, 
however profitably, to compare with data. Issues with such a prescription
include what average flux to employ to derive a representative dayside T/P profile,
how to incorporate longitudinal and latitudinal surface flows into the energy budget,
non-equilibrium chemistry\cite{non_chem}, photochemistry, and day-night differences when interested 
in total phase curves\cite{burrows_2008_close-in,budaj_coupling,spiegel_hatp7_tres2}.  
Nevertheless, such simple models are still commensurate with the 
information content of the extant observations.

The various quantities and topics that influence secondary eclipse spectra and 
have exercised the community include 1) the presence or absence of an extra 
absorber of currently unknown origin in the upper atmosphere that could 
heat those regions, at times producing thermal inversions over a restricted
pressure range\cite{burrows_2007_209,knutson_209_inver_2008}; 2) the temperatures 
and temperature profiles of the atmosphere; 3) the phase shifts from the orbital 
ephemeris of the light curves at various wavelengths and spectral bands due to zonal 
winds that redistribute heat\cite{burrows_rauscher}; 4) the compositions and 
elemental abundances of the atmospheres; 5) the presence of hazes and clouds; 
6) the day/night flux contrast; 7) Doppler signatures of atmospheric motions;
8) reflection albedos \cite{hd209_albedo_rowe}; and 9) the presence and role of evaporative planetary 
mass loss. I mention these challenges only to indicate the range of complex problems 
to be addressed, but will focus in this paper on only the simplest of approaches taken
to extract information from secondary-eclipse data.  

 
\begin{figure} 
\null\hskip-0.5cm
\begin{center}
\includegraphics[height=0.30\textheight,angle=-90]{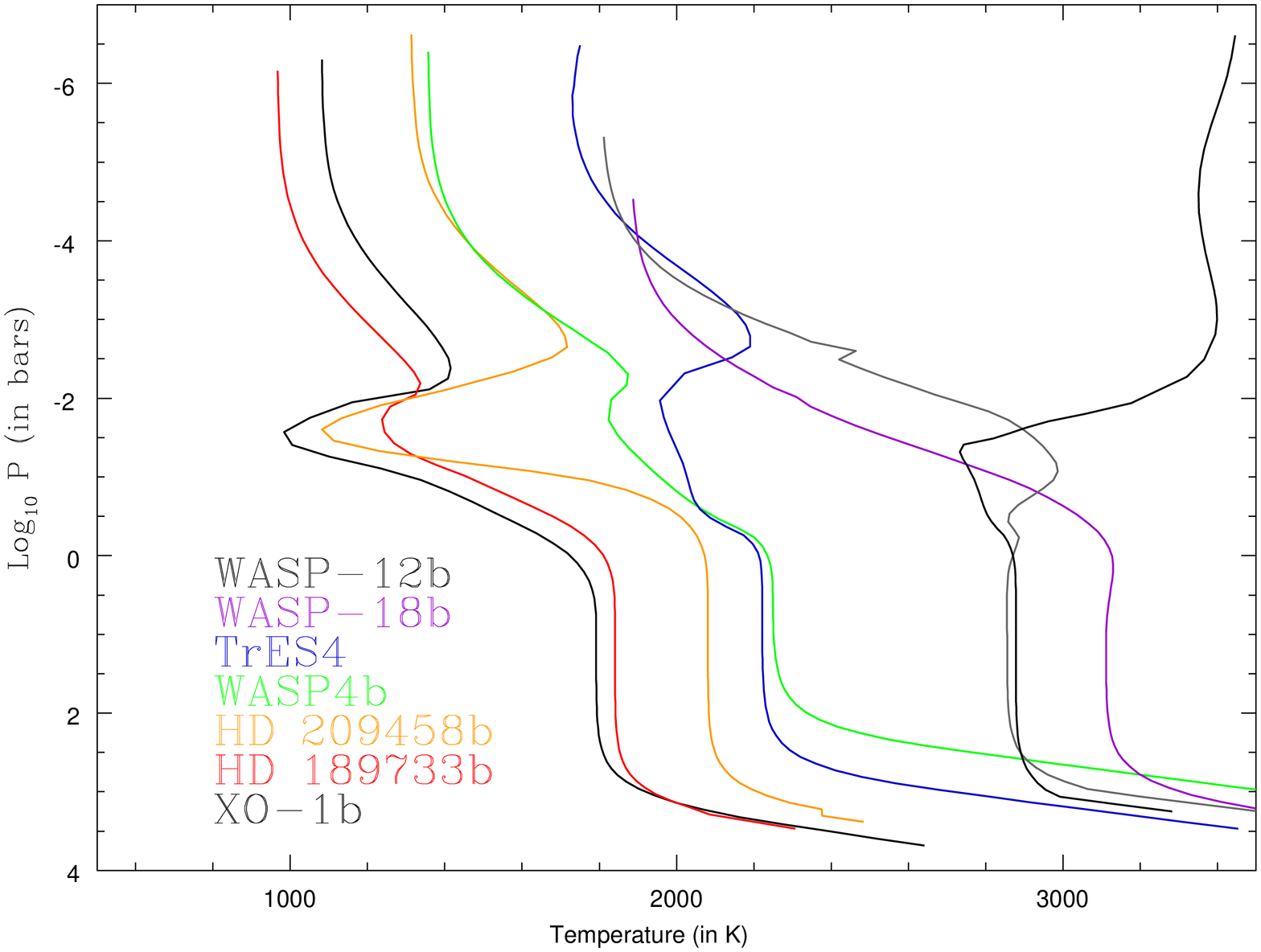}
\end{center}
\null\hskip-0.0cm
\begin{center}
\includegraphics[height=0.30\textheight,angle=-90]{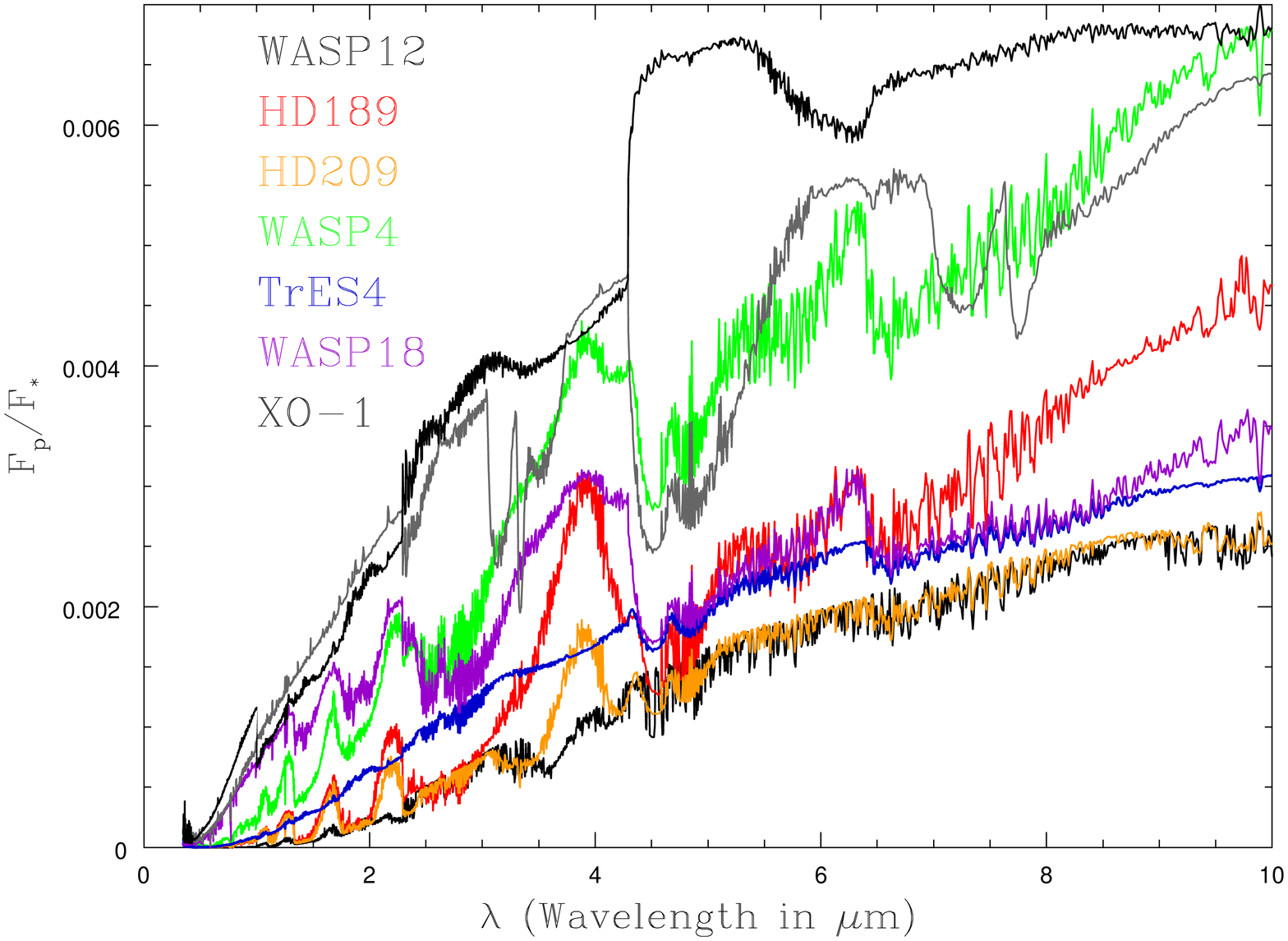}
\end{center}
\caption{{\bf Top:} The top figure portrays model temperature-pressure curves for a collection of transiting giant exoplanets.
These models were constrcuted in an attempt to fit respective Spitzer/IRAC data at secondary eclipse and demonstrate the span
of temperatures expected in giant exoplanet atmospheres.  This span reflects, among other things, the range of sub-stellar
fluxes at these given planets, as well as the extra heating of the upper atmosphere by an absorber in the optical that, at times,
has been invoked to explain Spitzer/IRAC data, in particular at 5.8 microns. Note that the XO-1b model is the black line at the left,
while the black line at the right is for WASP-12b with an inversion ($\kappa = 0.1$ cm$^2$ g$^{-1}$),
with P$_n$ = 0.1, and in chemical and radiative equilibrium at solar elemental abundances.  The gray curve is also a model for WASP-12b,
but without an inversion, depleted in water by a factor of 100, and enhanced in CH$_4$ and CO to uniform
fractional abundances of $2\times 10^{-4}$.  Each model was used to address the WASP-12b IRAC and near-infrared
secondary eclipse data.  {\bf Bottom:} The planet/star flux ratios from 0.4 to 10 microns
for the models on the top. The ``predicted" range in values, even for a class of solely giants, is very wide.
Note also that comparison between the two WASP-12b models (black and gray) is a cautionary tale
against relying too heavily on error-prone photometry to characterize exoplanet atmospheres, and a clarion call for accurate
spectra over a wide wavelength range.  See text for a discussion.
\label{fig3}}
\end{figure}

A few conceptual points are worth noting in passing:  1) An atmosphere calculation with
external incident flux will automatically generate a reflection albedo and
is not extra physics.  2) For a given elemental ratio set, the metallicity 
dependence of the emergent spectrum is quite weak.  Most relevant species (such as water)
have one ``metal" and incident and emergent integral fluxes must almost balance.
3) The difference between incident and emergent total fluxes is due to the 
true effective temperature ($T_{\rm eff}$), which for giant exoplanets
of Gigayear ages ranges from 50 to 500 K, and results in a very small contribution
to the emergent flux for a strongly irradiated planet. $T_{\rm eff}$ is important only when the
stellar irradiation flux is small, and this obtains only for wide-separation planets.  
4) The so-called equilibrium temperature, $T_{\rm eq}$,  is defined as the 
surface black-body temperature for which the incident stellar flux is balanced 
and is given by 
\begin{equation}
T_{\rm eq} = T_{*} \left(\frac{R_{*}}{a}\right)^{1/2}\left(f(1-A_B)\right)^{1/4}\, ,
\label{bondeff}
\end {equation}
where $T_*$ is the stellar effective temperature, $f$ is the heat redistribution 
factor\footnote{$f$ = 1/4 for isotropic models.}, and $A_B$ is the Bond albedo\cite{burrows_rmp_2001,sudarsky_albedo}.
While providing a measure of the mean temperature achieved in a planet's atmosphere,
assuming this can be used as the inner boundary condition $T_{\rm eff}$ has introduced 
quite a lot of confusion.  Very different T/P profiles can yield the same total flux,
but very different flux spectra.  Figures \ref{fig4_supplement} in the Supplement show two models
with the same emergent flux, and, hence, $T_{\rm eq}$.  One consistently incorporates
stellar irradiation, while the other puts a flux with $T_{\rm eff} = T_{\rm eq}$ at the base 
of the atmosphere.  Both are in radiative and chemical equilibrium.  As these figures demonstrate, 
despite the fact that the emergent fluxes are the same, the corresponding T/P 
profiles are hugely different and the flux densities at a given wavelength can be 
off by factors of 2$-$4!  Irradiated atmospheres are different from isolated atmospheres.  

Lastly, 5) if an atmosphere is in fact isothermal,
there must be an extra absorber in the optical at altitude.  Even under irradiation, the temperature
gradient must otherwise be negative from base to height, with characteristic temperature changes
of $\sim$500$-$1500 K for close-in giant exoplanets.  Hence, inferences of isothermality
are not as content-neutral as is often implied\cite{wasp12_crossfield}.

One can derive an average temperature profile in a radiative-equilibrium exoplanet atmosphere
under stellar irradiation by generalizing the classical Milne atmosphere\cite{hubeny_2003,guillot_2010}.
One obtains:
\begin{equation}
T^4 = \frac{3}{4}\, T_{\rm eff}^4\, \frac{\kappa_J}{\kappa_B}
\left[ \tau_R + \frac{1}{\sqrt{3}} \right] +
\frac{\kappa_J}{\kappa_B}\, W\, T_\ast^4\, ,
\end{equation}
where $W$ is the dilution factor, $(R_*/a)^2$, $\tau_R$ is the Rossleand depth, $\kappa_J$ is the photon energy-density weighted opacity,
$\kappa_B$ is the corresponding local black-body-weighted opacity, and we have used the Eddington 
approximation for the angular moments. For an isolated atmosphere, $\frac{\kappa_J}{\kappa_B}$ is 
close to one, but for an irradiated atmosphere $\kappa_J$ and $\kappa_B$ can differ appreciably.  
The former at altitude is dominated by the stellar spectrum, while the latter 
reflects the local atmospheric black-body spectral distribution. If this difference is an 
interesting function of altitude, an inversion can result\cite{hubeny_2003}.
We note that $T_{\rm eff}$ is generally small for close-in hot Jupiters. In this case, the temperatures
at depth are determined by the second term, which yields something like eq. (\ref{bondeff}). In reality,
gas giants are convective at high optical depths ($\sim$100-1000) and the T/P profile becomes
an adiabat.  Otherwise, it would be flat.

Representative theoretical average day-side temperature-pressure profiles for a subset of transiting gas giants
are given on the top panel of Figure \ref{fig3}.  These are provided to communicate the range of atmospheric
temperatures encountered for hot Jupiters and the matching to adiabats at depth above $\sim$100 bars.
The atmospheres of close-in giants can vary in temperature, depending upon $W$ and $T_*$, by $\sim$1000$-$2000 K.
Importantly, the difference in upper atmosphere temperatures between ``inverted" and non-inverted
situations can be $\sim$1000$-$1500 K, a huge difference that can translate into flux spectrum
differences of factors of $\sim$2$-$4 for ostensibly the same object. This is depicted on the bottom panel of Figure \ref{fig3}, which
provides the corresponding planet/star flux ratios versus wavelength.  Among this set of exoplanets, the theoretical 
flux ratios vary at a given wavelength by an order of magnitude. Moreover, as the comparison between 1) the model 
for WASP-12b with solar abundances, chemical equilibrium, and an extra upper atmosphere absorber in the optical (upper black)
and 2) the model for WASP-12b with enhanced CH$_4$ and CO, depleted H$_2$O, and no inversion (gray) attest, mid-infrared planetary 
spectra can vary significantly for the same stellar irradiation regime and gravity.  Figure \ref{fig3}, together with Figure \ref{fig1},
demonstrate the great diagnostic potential of multi-frequency spectra to extract compositions.  One can 
also determine the presence or absence of extra heating by enhanced absorption of stellar light that leads to 
inversions, but also hotter upper atmospheres and elevated fluxes of features formed in the heated zone.  
The pronounced bump at $\sim$4.5$-$5.5 microns on the inverted spectrum for WASP-12b is due to water in emission and the fact
that this band forms where the corresponding temperature profile has a positive (inverted) slope\footnote{Emission features
won't always be seen when the extra absorber is active $-$ this depends on where in the atmosphere the band is ``formed."}.

Though inversions have been inferred from enhanced {\it Spitzer} IRAC band fluxes (in particular at 5.8 microns), 
the nature of the absorber is still unknown.  It is suggested that TiO could do it, but there are good 
reasons to believe this compound would be rained out to depth by various cold traps\cite{spiegel_tio}. 
There may be a photochemical hazes with the right optical absorbing properties, but this has not been 
demonstrated. Still, it is tantalizing to hypothesize that the haze inferred by Pont et al.\cite{pont}
in the atmosphere of HD 189733b and that inferred by Deming et al.\cite{deming_haze} in the atmosphere of 
HD 209458b might in some way be implicated, or at least be of similar composition.

Though the interpretative and diagnostic promise of good spectra is suggested in Figure \ref{fig3}, the current 
reality is depicted in Figures \ref{fig5_supplement} (in the Supplement).  Here, I plot representative measured planet/star flux ratios
for 17 transiting giant exoplanets.  Most of the data are {\it Spitzer} IRAC photometry in four bands,
while some of the data are from the ground and the Hubble Space Telescope.  For HD 189733b, we have {\it Spitzer}/IRS spectra from $\sim$5 
to 15 microns as well\cite{grillmair}.  To keep the plots from being any more cluttered, 
error bars for only a few measurements are shown.  The quoted 1-$\sigma$ error bars generally range from $\sim$10\% to 30\%.
In an attempt to divide out universal expectations and to focus on what may distinguish one planet from another, 
I have normalized the planet/star flux ratio with the corresponding black-body value\footnote{Dividing by the factor
$\frac{f^{bb}_p}{f^{bb}_*}(\lambda) = \left(\frac{R_p}{R_*}\right)^2\left(\frac{e^{\frac{hc}{\lambda kT_*}} - 1}{e^{\frac{hc}{\lambda kT_p}}-1}\right)\, ,$
where $T_p$ has been set equal to $T_*\sqrt{\frac{R_*}{2a}}$.}

First, we see from Figures \ref{fig5_supplement} that the normalized ratio is rather flat over a broad range of wavelengths 
and close to one, perhaps a bit higher. However, the mean level could just reflect the crudeness of 
the $T_p$ employed for the comparison. We see undulations, but they have little information content, aside from the possible 
suggestion of enhanced or reduced flux in particular broad spectral regions. The IRS data near 
6.2 microns for HD 189733b do imply the presence of water, but what is the 
feature near 12.5 microns? There is a systematic increase in the ratios to shorter wavelengths, 
and this is probably real. As supplementary figures \ref{fig4_supplement} imply,  
fluxes from irradiated planets are expected to be mostly in the near infrared.  

The comparison of Figures \ref{fig3} and \ref{fig5_supplement} starkly emphasizes that we have a long way to go
before comparative exoplanetology becomes a richly diagnostic science.  At times,
data such as are depicted in Figure \ref{fig5_supplement} have been used to find temperatures, compositions, 
albedos, inversions, carbon-to-oxygen ratios \cite{wasp12_nikku}, metallicities, and day-night heat redistribution factors, etc.  Clearly,
these data, and the still primitive state of exoplanet atmosphere theory, do not justify attempts 
to constrain such quantities simultaneously, or perhaps at all.  Until high-quality transit and emission spectra
across a wide range of wavelengths are routinely available, only the most primitive
and conservative conclusions will be justified.  I reiterate that 
the data in Figures \ref{fig5_supplement} are for giant exoplanets. Smaller Neptunes, 
super-Earths, and terrestrial planets around similar stars will be much more difficult targets.  

\section{Systematic Uncertainties in the Data and Theory}
\label{uncertain_data}

Theorists and observers alike, anxious to extract all the conclusions they can 
from this first generation of measurements of exoplanet atmospheres,
have tended to overinterpret them.  A comparison between Figures \ref{fig3} and \ref{fig5_supplement} 
is a sober indication of the current limitations of the science. 
The telescopes being used were not designed with exoplanets in mind.  For example, {\it Spitzer} was designed
for photometry at the $\sim$1\% level, yet it is being used (however heroically) to obtain numbers
at the $\sim$0.1$-$0.01\% level.  Generally, the space-based and ground-based data have limited 
signal-to-noise, the systematic effects/errors are variously and imperfectly corrected for, there frequently
is no absolute calibration across disparate wavelength regions, stellar spots are difficult 
to account for, and corrections for the Earth's atmosphere for ground-based observations
have been problematic. Data for the same object at the same wavelength, but taken by different teams, have varied
by up to factors of $\sim$2, and such a factor can completely alter the conclusions drawn about 
abundances, C/O ratios, inversions, etc.  

Given this list of limitations, one should be highly sceptical of extraordinary claims based on imperfect
data with low intrinsic information content. Many published model fits have been highly underconstrained.
This is all the more important given the gross imperfections in current exoplanet atmosphere theory. 
With a few photometric points, one can not simultaneously determine with any confidence, 
or credibly incorporate into the fitting procedures, chemical and elemental abundances, wind dynamics, 
longitudinal heat redistribution, thermal profiles, albedos, the potential influence of hazes and clouds,
non-equilibrium chemistry and photochemistry, and magnetic fields. Furthermore, the opacities for many
chemical species are only imperfectly known, convection at depth is frequently handled with a mixing-length
approach, and emissions over a planetary hemisphere are never calculated with correct, multi-dimensional 
radiative transfer.  Moreover, most of the current generation of 3D general circulation models (GCMs) 
filter out sound waves, but derive transonic flows speeds with Mach numbers at and above one without 
a means to handle shock waves.  Many of these codes have also inherited from Earth GCM practice 
various ad hoc ``Rayleigh drag" and hyperdiffusivity terms with arbitrary coefficients calibrated on the Earth
that compromise the wind dynamics on strongly irradiated gas giants, even if magnetic torques are sub-dominant. 
Importantly, GCMs were configured to look at winds and pressures, not spectral emissions, highlighting the
mismatch between the traditional goals of planetary and Earth scientists and exoplanet astronomers.

At times, basic atmosphere practice has been shunted aside in attempts to retrieve thermal
and compositional information from a few (though precious) data points. 
Examples are 1) using unphysical, parametrized T/P profiles and arbitrary compositions, 
while not addressing local energy and chemical balance; 2) using 1D averaged models for what is a 3D planet;
3) using $T_{\rm eq}$ as if it were a real physical quantity of relevance to spectra; 
4) defining and deriving a reflection albedo when the planet is mostly emitting thermally; or 5) fitting 
photometric points with $T_{\rm eq}$ and a Bond albedo. Such approaches might seem right-sized 
to the data at hand, but are likely to generate an erroneous sense of confidence in the conclusions derived.
For example, it is long been known that small errors in $\Delta$T can translate into 
large spectral flux errors, even if the total reprocessed emitted flux is ostensibly addressed.

\section{The Future}
\label{training}

Therefore, I suggest that once high-quality, well-calibrated, stable spectra 
across a broad range of wavelengths from the optical to the mid-infrared are 
finally available many conclusions reached recently about exoplanet atmospheres 
will be overturned.  The current interpretations and theories are just not
robust enough to survive intact into the future. 
However, despite the generally cautionary tone of much of this paper, I see an exciting future.
The past $\sim$20 years has been but a training period for a new generation 
of exoplanet scientists, forged by trial and error and educated in the new questions 
posed by exoplanets.  Its growing membership is testing its tools $-$ new  
technologies, concepts, theories, and techniques $-$ that will serve to establish 
a solid foundation for a true science of planets not tethered to the solar system.  
Informed by the latter, but optimized to address its unique challenges as a 
remote-sensing science, comparative planetology's youth is rapidly maturing. 

The near- and mid-term future of exoplanet atmosphere characterization
will include the James Webb Space Telescope (JWST)\cite{deming_JWST,shabram}, ground-based 
Extremely-Large/Giant-Segmented-Mirror Telescopes (ELTs/GSMTs)\cite{angel}, and 
perhaps dedicated Explorer, M-Class (e.g., EchO\cite{echo2}), or Probe-Class 
space missions. The continued creative use of existing ground-based telescopes 
is assured, and new high-contrast coronagraphic imaging programs now coming on 
line (such as GPI\cite{GPI} and SPHERE\cite{SPHERE}) show great promise. 
Importantly, there is the exciting possibility of putting a
coronagraph on WFIRST/AFTA \cite{spergel}. In the farther future, 
once a cost-effective plan can be articulated, a major dedicated space mission 
of exoplanetary atmosphere characterization, such as was envisioned with 
the TPFs and Darwin, should be possible.  Currently, giant planets and Neptunes pose the most
realistic targets, but terrestrial planets and the possibility of discerning signatures of life
are majors goal of many.  Soon, the spectra of terrestrial planet atmospheres around 
small M-dwarf stars may be within reach.  

Given this, it is clear that, for the field to remain vibrant and grow, it needs 
a heterogeneous and balanced program of ground-based and space-based facilities 
and programs.  If anything has been demonstrated by the first $\sim$20 years
of exoplanet research, it is that some of the best techniques for studying them
are unanticipated.  The transit technique for close-in planets has been a game-changer, but was not 
envisioned in previous planning documents. High-contrast imaging, only now coming of age, 
was to inaugurate the era of atmospheric characterization. It is also clear that large, expensive missions
are counterproductive until they are demanded by the science, in fact until the science indicates
that further progress demands them. Precursor technologies for such missions should certainly be 
pursued and allowed to compete.  But overlarge and expensive missions without the 
requisite credibility and technological heritage in place can fatally squeeze the smaller programs 
that have proven so fruitful. This implies that an international Roadmap should be crafted 
for exoplanet's next $\sim$20 years. Its guiding principle should be a balanced 
approach of small, medium, and large initiatives that encourages flexibility and scientific return, 
and does not presume (or proscribe) a specific future.  The clear goal is to understand in rich detail
the planets that we now know exist in profusion in the galaxy and Universe.
One is only left to ask: Are we ready to assume the challenge?

\begin{acknowledgments}
The author would like to acknowledge support in part under NASA ATP grant NNX07AG80G,
HST grants HST-GO-12181.04-A, HST-GO-12314.03-A, HST-GO-12473.06-A, and HST-GO-12550.02,
and JPL/Spitzer Agreements 1417122, 1348668, 1371432, 1377197, and 1439064.
A diverse suite of exoplanetary spectral and evolutionary models for a range of masses and compositions
is available at {http://www.astro.princeton.edu/$^{\sim}$burrows} and from the author upon request.
\end{acknowledgments}

\newcounter{firstbib}

\renewcommand{\refname}{Reference 1}

\newpage

\centerline{{\bf Figure Supplement to ``Spectra as Windows into Exoplanet Atmospheres"}}

\begin{figure*} [h!]
\null\hskip-1.0cm
\begin{center}
\includegraphics[height=0.33\textheight,angle=-90]{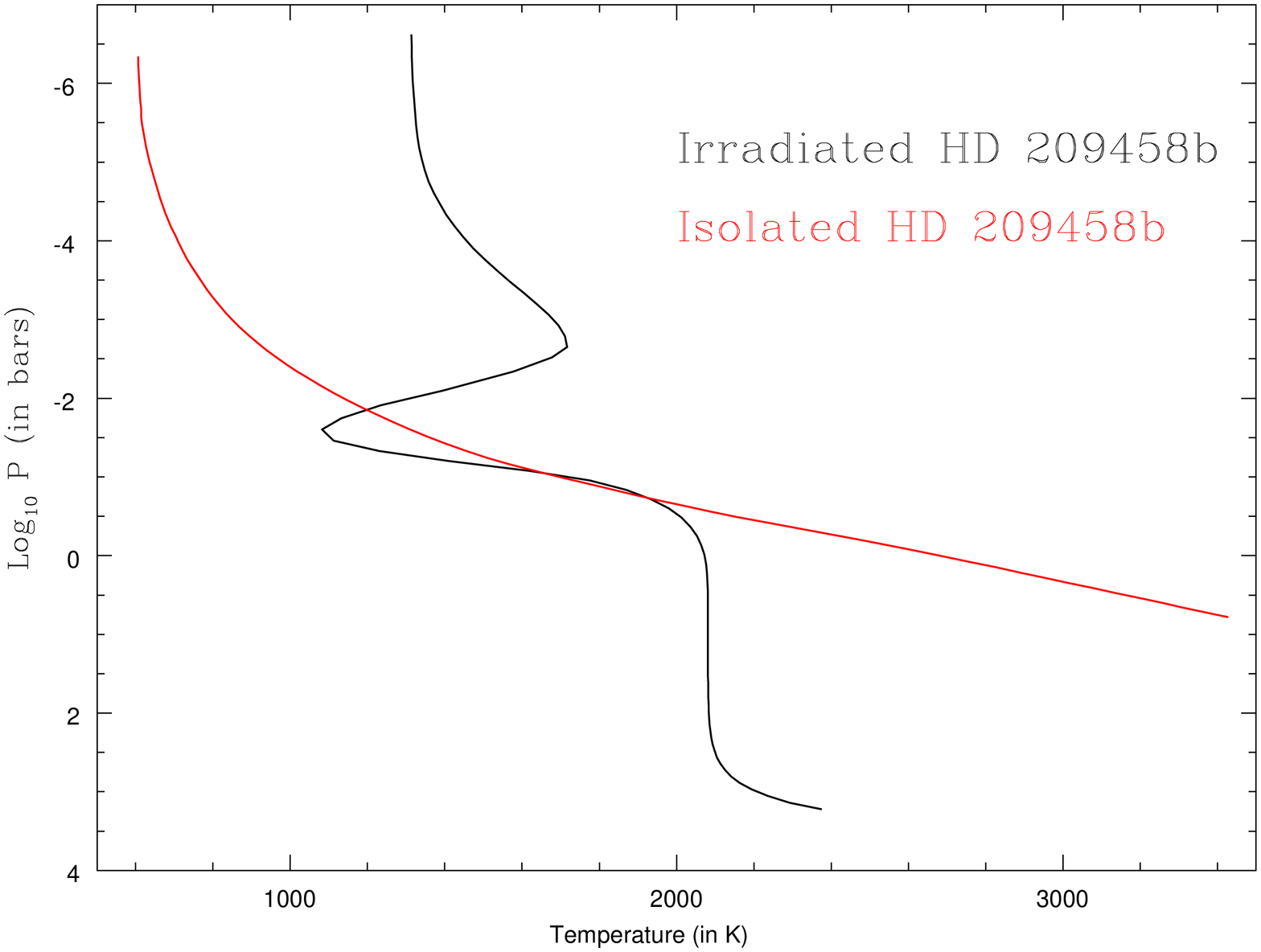}
\includegraphics[height=0.33\textheight,angle=-90]{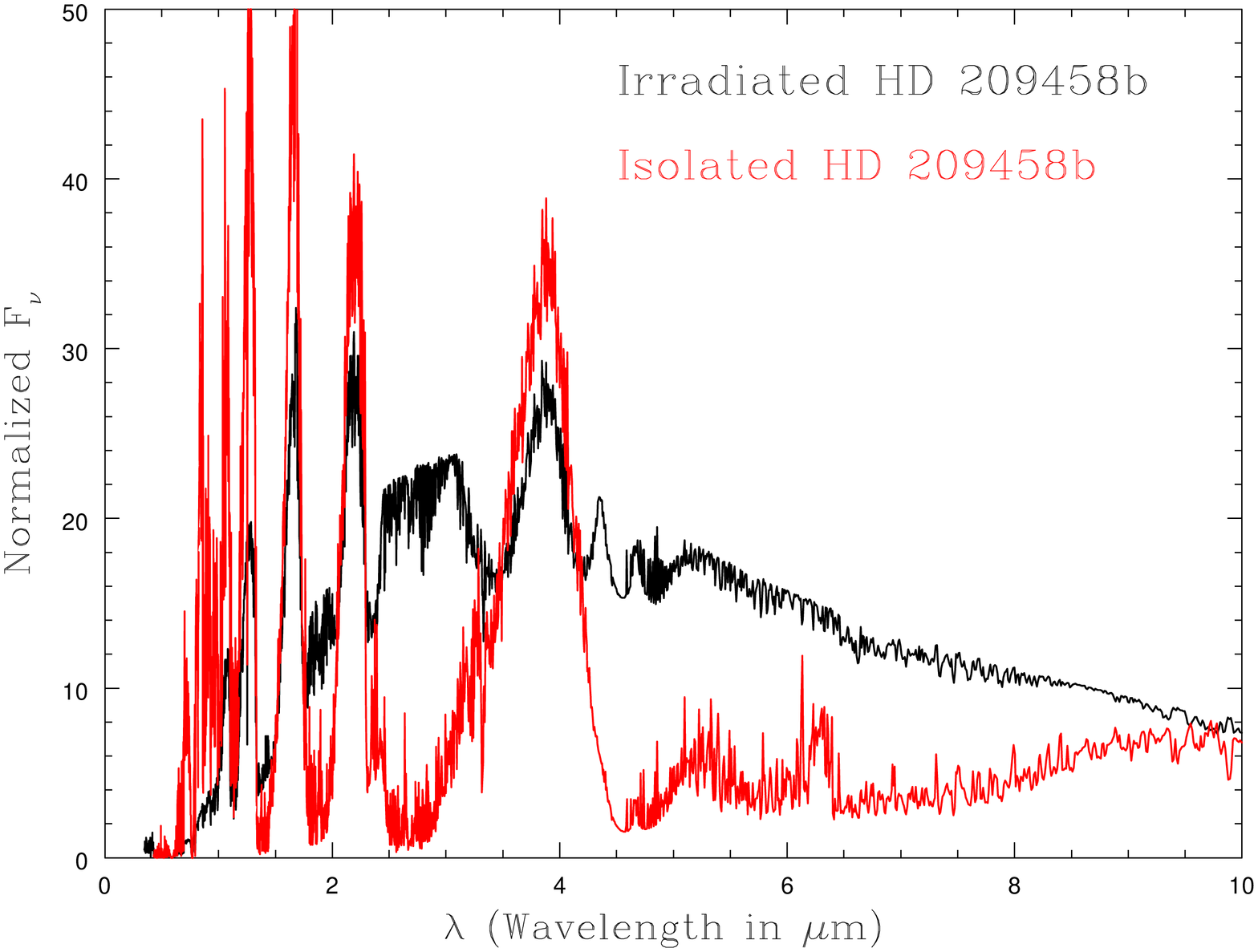}
\end{center}
\null\hskip-0.0cm
\caption{{\bf Left:} Shown are the temperature-pressure profiles for two models of HD 209458b.  The black curve was
generated including the stellar irradiation flux at the orbital distance of the planet
and a token effective temperature (T$_{eff}$) of 200 K at the base.  Note that T$_{eff}$ for such
a model reflects the {\it net} flux, not the emergent flux.  The red curve is for an isolated model with
roughly the same total emergent flux at an effective temperature T$_{eff}$ of 1700 K.
Despite having the same emergent flux, these temperature-pressure profiles are profoundly different. {\bf Right:} The bottom panel
depicts the corresponding normalized spectra, F$_{\nu}$, versus wavelength (in microns).  These spectra are vastly different, though
the total emergent fluxes are the same, and demonstrate that one cannot assume that an equal emergent flux constraint
will translate into useful spectra or colors.  They also demonstrate that one must be careful when quoting an effective temperature,
and not confute T$_{eff}$ with an ``equilibrium temperature," T$_{eq}$.  See text for a discussion.
\label{fig4_supplement}}
\end{figure*}

\renewcommand{\refname}{Reference 2}

\begin{figure*}  [h!]
\null\hskip-1.0cm
\begin{center}
\includegraphics[height=0.35\textheight,angle=-90]{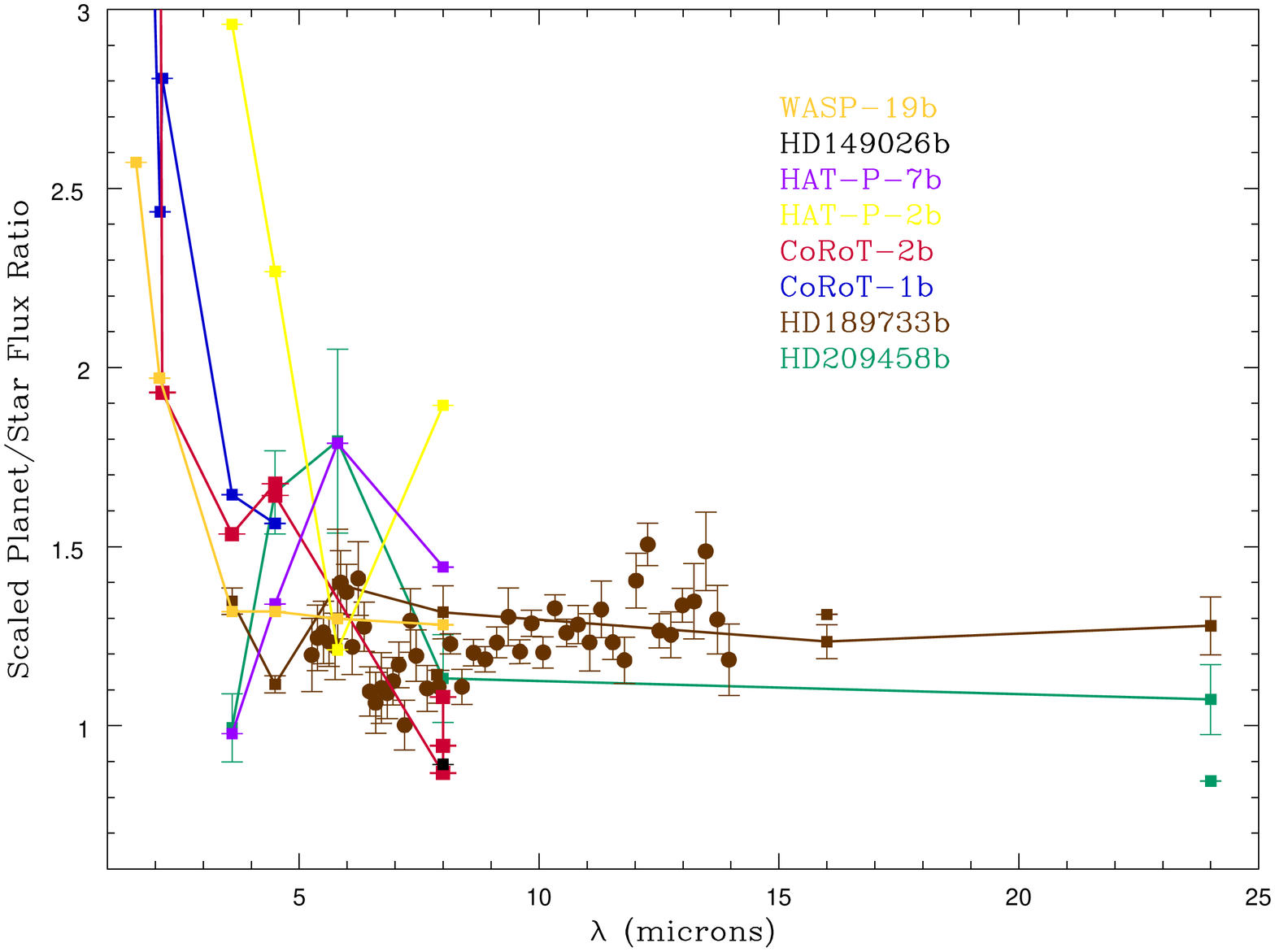}
\includegraphics[height=0.35\textheight,angle=-90]{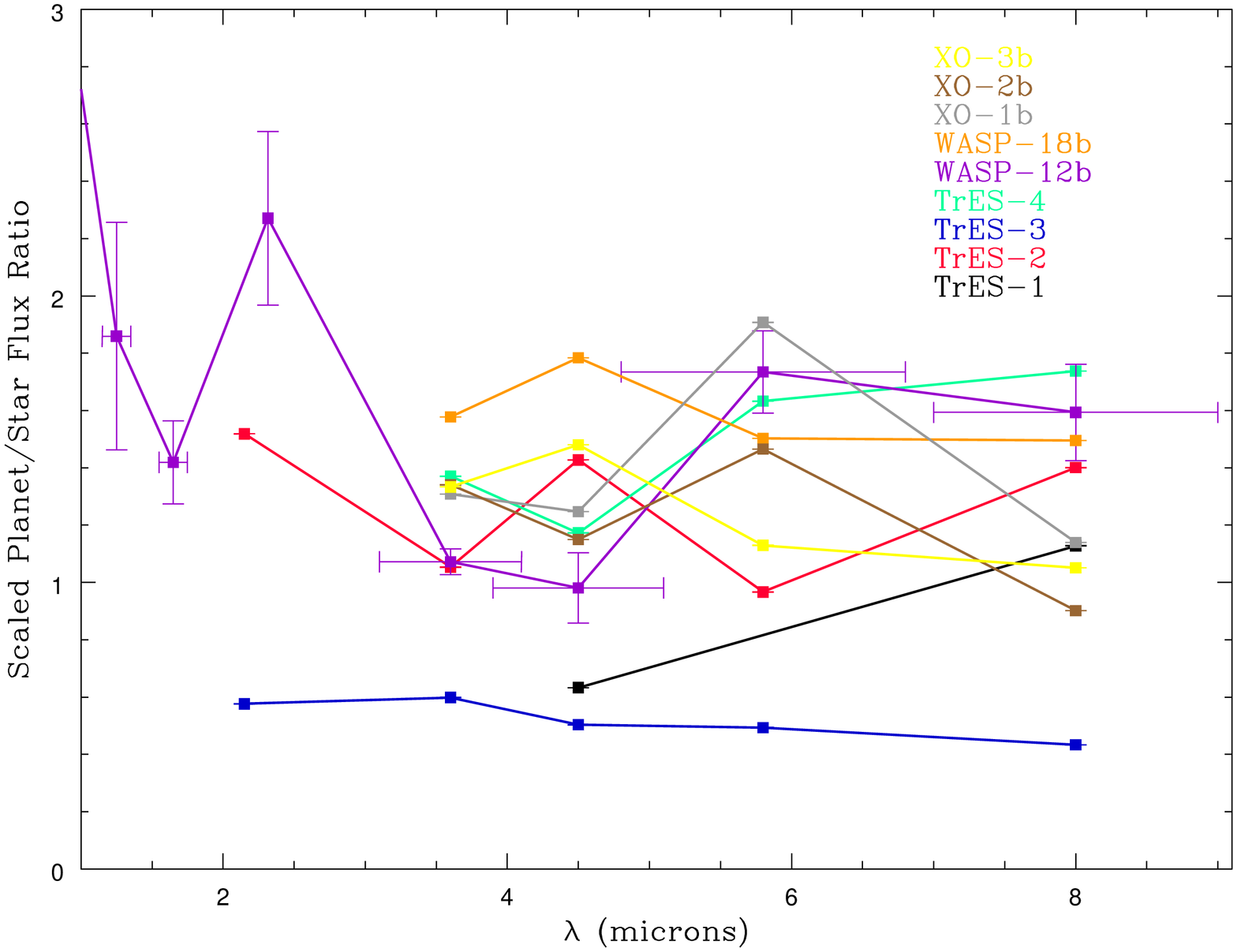}
\end{center}
\null\hskip-0.5cm
\caption{{\bf Left:} Planet/Star flux ratio data points at secondary eclipse for eight giant planets (WASP-19b, HD 149026b, HAT-P-7b,
HAT-P-2b, CoRoT-2b, CoRoT-1b, HD 189733b, and HD 209458b), normalized to the corresponding ratio
if both star and planet were black bodies at the corresponding measured stellar T$_{eff}$ = T$_*$
and zero-albedo equilibrium temperature, T$_{eq}$ $\left(=T_*\sqrt{\frac{R_*}{2a}}\right)$, respectively.  The lines connect points
for the same object.  Most of the data are Spitzer/IRAC points, but points at shorter wavelengths, where available,
are also included.  For HD 209458b and HD 189733b, points at 16 and/or 24 microns are also given, along with points (unconnected and
for comparison) derived from other reductions. To avoid further clutter, quoted error bars are given
only for the IRS spectrometer data for HD 189733b and the Spitzer data for HD 209458b.
{\bf Right:} The same as on the top, but for XO-3b, XO-2b, XO-1b, WASP-18b, WASP-12b, TrES-4, TrES-3, TrES-2, and TrES-1.
Error bars for only WASP-12b are given.  The normalization provided helps to rationalize the interpretation potential
of such photometric and low-resolution data and to facilitate planet-planet comparison.
The data were taken from \cite{tres4,tres3,tres3_croll,hd149026b,hd189_2012_new,hd189_agol_8micron,charb_189_irac,grillmair,
deming_16micron_189,tres1,knutson_209_inver_2008,hd209_deming_24,tres2,kipping_bakos_tres2,tres2_croll,hatp7_epoxi,
wasp12_croll,wasp12_cowan,wasp12_crossfield,wasp12_morales,wasp18_nymeyer,corot1_corot2_deming,
corot1_alonso,corot1_gillon,corot1_snellen,corot1_rogers,corot2_alonso,corot2_alonso_corr,corot2_snellen,
corot2_gillon,hatp2_lewis,xo1_machalek,xo2_machalek,xo3_machalek,anderson}. See text for a discussion.\label{fig5_supplement}}
\end{figure*}

\end{article}

\end{document}